# Controlling the Manifold of Polariton States Through Molecular Disorder

*Aleesha George, Trevor Geraghty, Zahra Kelsey, Soham Mukherjee, Gloria Davidova, Woojae Kim, Andrew J Musser*[*]

**Abstract**

Exciton polaritons, arising from the interaction of electronic transitions with confined electromagnetic fields, have emerged as a powerful tool to manipulate the properties of organic materials. However, standard experimental and theoretical approaches overlook the significant energetic disorder present in most materials now studied. Using the conjugated polymer P3HT as a model platform, we systematically tune the degree of energetic disorder and observe a corresponding redistribution of photonic character within the polariton manifold. Based on these subtle spectral features, we develop a more generalized approach to describe strong light-matter coupling in disordered systems that captures the key spectroscopic observables and provides a description of the rich manifold of states intermediate between bright and dark. Applied to a wide range of organic systems, our method challenges prevailing notions about ultrastrong coupling and whether it can be achieved with broad, disordered absorbers.

## 1. Introduction

Exciton-polaritons are quasiparticles that result from the strong interaction between light and matter, resulting in hybridization between well-defined photonic states – for instance in optical cavities, nanoparticle arrays, or nanostructured materials – and electronic excitations.[1–4] Though originally studied in highly ordered, cryogenic inorganic semiconductors, exciton-polaritons can also be readily observed at room temperature in a wide array of organic systems.[5,6] The foundational understanding of organic exciton-polaritons is derived from studies of materials characterized by minimal electronic disorder and narrow linewidths, such as cyanine dye J-aggregates.[5,7–12] However, more recently there has been a paradigm shift towards organic materials with markedly higher disorder. Broad molecular absorbers have been harnessed to explore a wide array of polaritonic phenomena, including triplet harvesting [13], charge transfer within photovoltaic devices, [14–16] long-range energy transport, [17–19] and condensation.[20,21] Despite this change in focus to disordered systems, there has been little explicit consideration of this material parameter and most analyses remain based on the models from well-ordered systems.

The conventional approach to strong coupling stems from the Tavis-Cummings (TC) model, which describes the interaction between $N$ degenerate two-level systems and a single field quantum (photon).[22] The Hamiltonian takes the form [23]:

$$H_{TC} = \sum_k^N \frac{1}{2}\hbar\omega\sigma_k^+\sigma_k^- + \hbar g\left(\sigma_k^+ a + a^\dagger \sigma_k^-\right) + \hbar v a^\dagger a \qquad (1)$$

A. George, T. Geraghty, Z. Kelsey, S. Mukherjee, G. Davidova, A.J. Musser
Department of Chemistry and Chemical Biology
Cornell University, Ithaca, New York 14853, USA
Email: ajm557@cornell.edu
W. Kim
Department of Chemistry
Yonsei University, Seoul 03722, Republic of Korea

where $k$ indexes the excitons collectively interacting with single-mode EM field of energy $\hbar v$. The resonant energy of the two-level system is given by $\hbar \omega$, $\sigma_+$ and $\sigma_-$ are exciton raising and lowering operators, $a^\dagger$ and $a$ are photonic creation and annihilation operators, and coupling constant $g$ parameterizes the light-matter interaction strength. In matrix form, this system can be readily block-diagonalized into $N+1$ eigenstates, of which only two are bright (i.e., mixed with the photon): the lower (LP) and upper polaritons (UP). The other $N-1$ states remain resonant with the original exciton and are optically dark.[24,25] Implicit in this picture is that the excitons exhibit only homogeneous broadening. To explicitly incorporate the effect of linewidth, an imaginary factor inversely proportional to the exciton or photon lifetimes is introduced. Omitting the unperturbed dark states for simplicity, the resulting $2 \times 2$ Hamiltonian describing the light-matter coupling is:

$$\widehat{H}_{TCS} = \begin{bmatrix} E_c(\theta) - i\Gamma_c & \Omega \\ \Omega & E_x - i\Gamma_x \end{bmatrix} \qquad (2)$$

where $E_c(\theta)$ captures the angular dispersion of the cavity mode, $E_x$ is the exciton energy, $\Gamma_c$ ($\Gamma_x$) is the photon (exciton) damping, and $\Omega = \hbar g$. This coupled-oscillator form permits ready extension to the strong-coupling of more complex optical spectra. Rather than a single optical transition, typical organic semiconductors are characterized by a progression of vibronic bands. These are generally incorporated as additional distinct excitons, each of which can separately couple to the photonic state. The result is a ladder of lower, middle, and upper polariton states, with each pair of polaritons separated by dark states.

Crucially, this model presupposes that each coupled transition reflects a set of degenerate states, but that is not consistent with the excitonic structure of organic materials.[26,27] The extension of this formalism to incorporate inhomogeneous broadening has been pursued in the context of both inorganic and organic semiconductors, based on idealized optical spectra. Unsurprisingly, a widespread conclusion is that exciton broadening gives rise to broadened dark states.[28] Multiple works have further underscored a major impact of such disorder on the nature of the coupled eigenstates, with significant state mixing induced within the dark manifold of the TC model and thus a substantially more complex polariton manifold.[29–32] Not only is mixing predicted between different material states, the presence of disorder additionally causes these dark states to mix with the photon mode.[29,33,34] The resulting subradiant or gray states are heavily mixed, with varying degrees of photonic character that mark them as distinctly intermediate between the canonical bright and dark states. Their emergence is suggested to open entirely new photophysical channels including more efficient long-range energy transport[35–37] and significantly extended excited-state lifetimes.[38]

Despite this solid theoretical understanding that the simple picture of TC is altered in the presence of disorder, its impact is scarcely considered experimentally. Only in one recent study was the functional influence of disorder directly investigated, through combining distinct carbon nanotubes in the strong-coupling regime.[38] This work proposed that the broad distribution of inter-nanotube couplings resulted in the formation of a broad manifold of gray states possessing varying degrees of photonic character. This redistribution of photon content extended the lifetime of each eigenstate, resulting in unexpectedly gradual and long-range energy transfer across the cavity. Though direct spectral evidence of these gray states remained elusive, this kinetic effect provides strong evidence for the disorder-induced expansion and mixing of the dark-state manifold. However, the standard approach for even much more disordered systems than the carbon nanotubes remains to expand the TC model of Eqn. (2) to encompass a minimal set of distinct vibronic excitons, regardless of how well resolved they are. This approach can generally capture the most prominent spectral features in angle-resolved reflectance, but it fails to account for the subtle spectral modulations that are frequently observed in the gap between the LP and UP, while in other cases it predicts middle polariton features which cannot be experimentally detected.[21,39–41]

There is thus a growing need for an experimentally driven approach to understand the role of disorder in polariton formation and bridge the gap between measurement and advanced theoretical models. Here, we explore strong coupling in a conjugated polymer with controllable disorder, P3HT, to systematically tune the redistribution of photon character through the band of intermediate states. Motivated by the observation of subtle spectral features that are widely overlooked, we demonstrate how a straightforward extension of the TC model—by representing inhomogeneous broadening through a dense array of Lorentzian oscillators—provides a much more accurate description of the optical behavior. At the same time, this intuitive approach built directly on measured spectra captures the same essential physics as advanced theoretical methods, providing a simple tool to accurately describe the rich manifold of states formed by strong coupling in disordered materials. Strikingly, our method provides an excellent description of the behavior of systems typically assigned to be in the ultra-strong coupling (USC) regime, without meeting any of the standard hallmarks of USC. These results suggest that broad absorbers may not, in fact, be a suitable platform to achieve the intriguing physics of USC, and that the threshold for this regime should be more carefully considered.

## 2. Experimental Methods

### 2.1. Materials

The chemical structures of the materials employed here are depicted in Figure 1a. The prototypical conjugated polymer P3HT (Mw=74000, 97.3% regioregular, Ossila) offers the scope to control the degree of interchain packing in thin films through processing conditions such as solvent selection, concentration, and temperature.[42,43] The resulting tunability of the P3HT absorption spectrum makes it ideal to explore disorder effects in the strong-coupling regime. Lemke dye (Aurora Analytics) offers a simple, featureless primary absorption band and has previously been reported to achieve USC in Fabry-Perot cavities.[44] TIPS-pentacene (Sigma Aldrich) and PTCDA (Sigma Aldrich) were selected to represent two extremes of spectral behavior: narrow, well-resolved vibronic peaks in TIPS-pentacene [45] versus broad, overlapping bands in PTCDA.[46] Gelatin (type B from Bovine, gel strength ~225 g Bloom, Sigma Aldrich), polymethylmethacrylate (PMMA, Mw= 120,000, Sigma Aldrich), and polystyrene (PS, Mw= 35,000, Sigma Aldrich) were incorporated as inactive spacer layers or matrices. Samples were prepared on ultra-flat quartz coated glass substrates (Ossila). All materials were used as received.

### 2.2. Sample preparation

We prepared three classes of P3HT structure characterized by the degree of order in the active layer, as determined from the vibronic structure evident in steady-state absorption spectra. To prepare crystalline films, we spin-coated a 2wt% solution of P3HT in dichlorobenzene at varying spin speeds (1000-6000rpm, Ossila L2001A3) for 60 seconds. We deposited semi-crystalline films from solutions of P3HT in chloroform using two methods: a 1 wt% solution spun at 1500rpm for 60 seconds, or a 0.5wt% solution spun at 1000rpm for 60 seconds followed by thermal annealing at 200 °C for 300 seconds. This condition was identified to replicate the partial annealing that can occur during the deposition of semitransparent top Ag mirrors on amorphous films to form optical microcavities (see below). Finally, we prepared amorphous films from the same 0.5wt% solution of P3HT in chloroform, spun at higher speeds than 1000rpm without annealing. We followed a similar procedure to prior reports to spin-coat films of Lemke dye.[44] We combined stock solutions of 20mg mL$^{-1}$ Lemke dye in toluene and 30mg mL$^{-1}$ PMMA in toluene to achieve Lemke dye:PMMA ratios of 1:0, 8:1, 6:1, 3:1, 2:1, 1:2, and 1:9. Solutions were spun at variable speed to achieve thicknesses of 120-150 nm. We prepared films of TIPS-pentacene dispersed in PS in a 4:1 ratio (30 mg mL$^{-1}$ TIPS-pentacene, 7.5 mg mL$^{-1}$ for PS in toluene), using spin speeds of 2000-5000 rpm to achieve 150-180 nm films. We deposited ~100 nm films of neat PTCDA using thermal evaporation (Angstrom Engineering) at 333 °C and a rate of 1 Å/second.

We attained strong coupling to these materials in Fabry-Perot optical microcavities based on the structures in Figure 1b-c. Following thermal evaporation of a 200 nm Ag bottom mirror, we deposited active layers as described above to bring the λ/2 cavity mode into resonance with the main absorption band. A semitransparent 30 nm Ag top mirror was evaporated directly onto the active layer, and all measurements were performed from this side. As needed, we adjusted the cavity thickness by incorporating additional spacer layers spin-coated from a gelatin solution (10 mg mL$^{-1}$ in water). In the case of amorphous P3HT, we observed that even with protective gelatin layers the deposition of the top mirror partially annealed the film, resulting in increased prominence of the features associated with semi-crystalline films (SI Figure 1a-b).

### 2.3. Characterization

We recorded UV-vis absorption spectra of all the films on a home-built absorption spectrometer using a Xe plasma white-light source (LDLS, Hamamatsu) and AvaSpec-Mini4096 detector (Avantes). To characterize the angular dispersion in microcavities, we incorporated this source and detector into a home-built, fiber-coupled goniometer with two rotating arms.

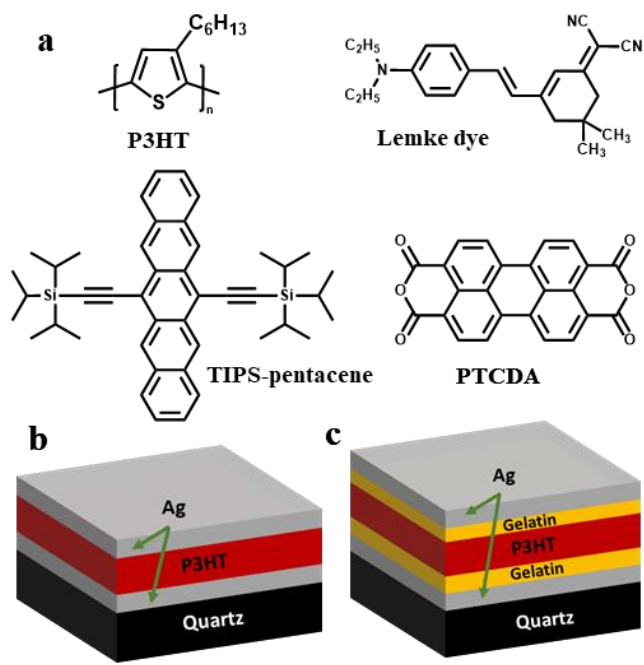

**Figure 1.** (a) Chemical structure of the molecules used in this study. (b) Architecture of P3HT cavity. (c) The cavity configuration employed to create amorphous and concentration dependent crystalline P3HT microcavities, integrating a gelatin layer to account for the overall thickness.

### 3. Results and Discussion

As shown in Figure 2a-c, we can readily distinguish the degree of interchain packing order in our three types of P3HT film from the optical absorption spectra. The crystalline and semi-crystalline films exhibit clear vibronic peaks at 2.05 eV, 2.23 eV, and 2.37 eV (red arrows). The increased prominence of the lower-energy bands is characteristic of H-type interactions induced by inter-chain π-π stacking,[47,48] associated with higher crystallinity. On the other hand, increasing intensity at the 2.37 eV peak is characteristic of a higher content of amorphous regions in the film.[47] Nonetheless, even in our most amorphous films the same progression of vibronic bands can be discerned at the same energies. That is, through our series of films we maintain an identical set of exciton transitions. Controlling the degree of order simply increases the linewidth of each band and slightly redistributes the absorption intensity.

The angle-dependent reflectivity of the corresponding microcavities is shown in Figure 2d-f. Across all three cavity types, we observe the same pair of clear angle-dependent minima in the reflectivity: a narrow one at low energies and a broad one at high energies. Both bands occur at energies distinct from the bare P3HT transitions. Moreover, these features exhibit angle dependence which anti-crosses the main exciton bands, revealing a dispersion pattern that is distinct from that of a bare photonic mode (e.g., in a cavity with non-absorbing spacer, see SI Figure 2). These characteristics are hallmarks of the strong-coupling regime, and we assign these prominent transitions as lower and upper polaritons (LP, UP). From the minimum separation between these bands, we extract a Rabi splitting of 1.02 eV in the crystalline cavity, roughly 42% of the P3HT transition energy. Comparably large Rabi splittings of 0.84 eV in semi-crystalline and 0.87 eV in amorphous cavities, that is, ~35% of the exciton energy, suggest all structures surpass the threshold for strong coupling and even the USC regime. [49,50] This result is consistent with the prior report of semi-crystalline P3HT microcavities, [51] but we return to the assignment of USC in Section 3.5 below.

Here, we highlight that the determination of USC and extraction of the Rabi splitting is complicated by the presence of distinct states observed between LP and UP. The crystalline cavity exhibits two additional well-defined bands, as indicated with arrows at 2.08 eV and 2.31 eV in the 10° spectral cut of Figure 2d. These features appear between the primary exciton bands over the full angular range and thus exhibit the same anti-crossing behavior as LP and UP. Within the semi-crystalline and amorphous cavities, the positions of the LP and UP bands remain consistent with those observed in the crystalline cavities. However, revealing differences emerge in this intermediate spectral region. Unlike the crystalline cavities where two distinct bands with varying degrees of visibility are observed, the less structured cavities each exhibit a single, relatively subtle modulation in the reflectivity at 2.1 eV. Similar intermediate modes, positioned between much stronger LP and UP bands, can frequently be distinguished in the reflectivity spectra of other broad absorbers operating in the strong-coupling regime, but they are typically overlooked due to their limited visibility[40,41,51] However, these modes originate from the influence of strong light-matter coupling, and thus they provide a useful and under-utilized spectroscopic handle to provide deeper insight into the microcavity energetic structure.

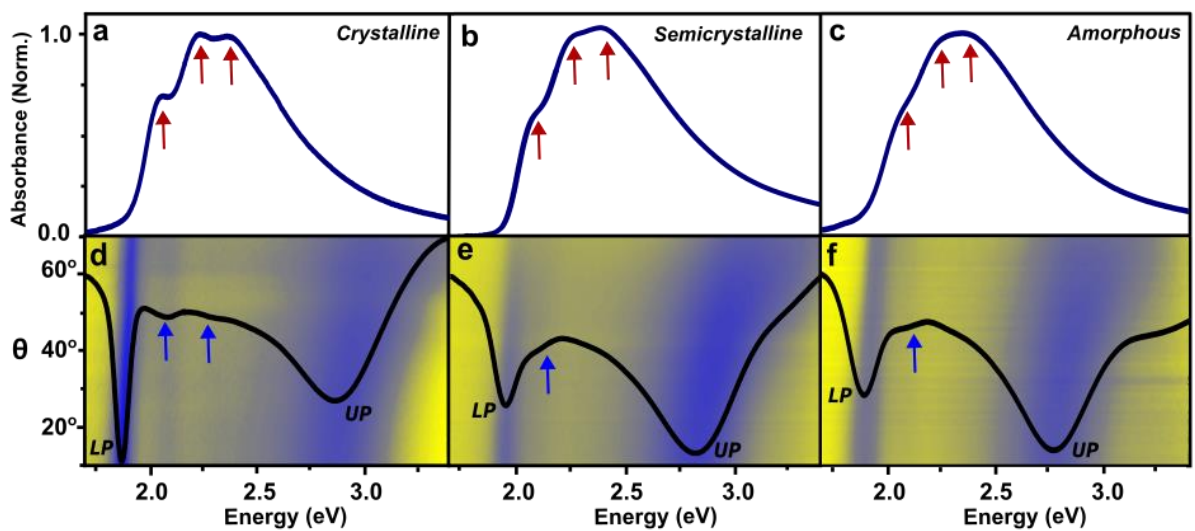

**Figure 2.** Steady state characterization of bare films and cavity. Normalized absorption spectra of (a) crystalline, (b) semi-crystalline, (c) amorphous P3HT thin films. Red arrows highlight the vibronic peaks used to characterize film

type. (d-f) Angle-dependent reflectivity of microcavities corresponding to the films above. Black solid lines show the reflectivity at an angle of 10°. Blue arrows highlight the intermediate bands between prominent LP and UP.

### 3.1. Optical modelling

*3.1.1. Conventional treatment*

The standard application of the TC Hamiltonian to organic films involves decomposing the linear absorption spectrum into a minimal series of peaks to capture the vibronic progression. In the case of crystalline P3HT, we used a three-Lorentzian fitting approach to describe the steady-state absorption, with the resulting exciton peaks and homogeneous linewidths incorporated into a standard coupled-oscillator model following Equation 2. Following diagonalization, the imaginary part of the resulting eigenstates reflects their linewidth, while the photon fractions (Hopfield coefficients) determine the strength of the angle-dependent optical response. Thus, this conventional coupled-oscillator model output is readily converted into a predicted reflectivity dispersion, as plotted in Figure 3a. Evidently this approach captures the approximate positions of LP, UP, and the intermediate bands in the crystalline P3HT microcavity, but it strongly overestimates the intensity of the middle bands compared to the experimental observations in Figure 2d.

This disparity stems from the coarse approximation built into the model that exciton damping is the sole source of broadening, rather than accounting for the underlying disorder these lineshapes represent. That is, such coupled oscillator models treat the excitonic transitions as homogeneously rather than inhomogeneously broadened. In a system like our P3HT microcavities, where the same three transitions are always present and films are chiefly distinguished by their linewidth, this three-oscillator model struggles to accurately capture the intensity variation of the middle bands in spite of the accurate description of bright LP and UP (see SI Figure 3a-b). This model can nicely describe systems with well-resolved absorption peaks such as TIPS-pentacene (Figure 3b-c), capturing the full experimental dispersion including prominent middle polaritons between 1.95 eV and 2.2 eV. Though there are still quantitative disparities, such as an overestimate of the band intensity at 2.25 eV, the agreement for TIPS-pentacene is clearly better than for P3HT. We argue, then, that this minimal coupled-oscillator picture falls short specifically for organic materials dominated by inhomogeneous broadening. Given the widespread and growing use of disordered molecular materials in polaritonic studies, this limitation emphasizes the need to develop a more general and experimentally driven approach to describe the states formed in the strong-coupling regime.

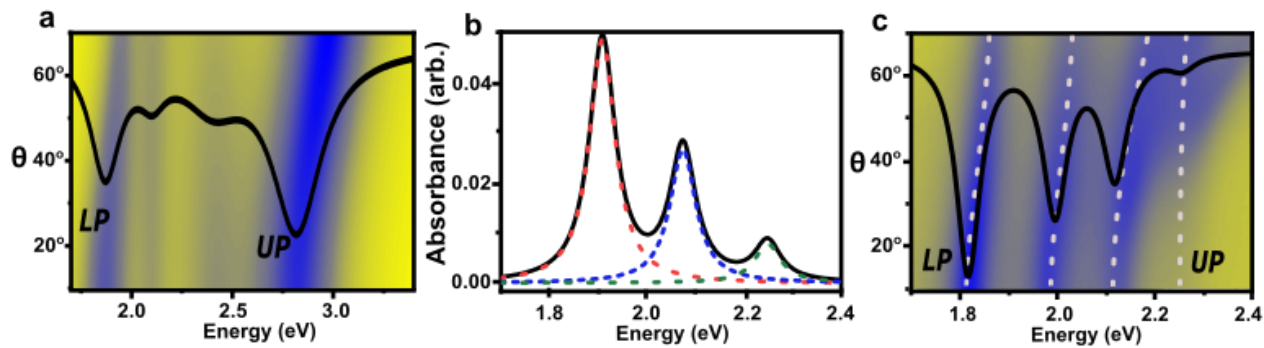

**Figure 3.** (a) Reflectivity map of crystalline P3HT based on standard coupled oscillator fit overlaid with spectral cut at 10°. (b) Steady-state absorption spectrum (black lines) and coupled oscillator fit (dashed lines) for TIPS-pentacene dispersed in a polymer film. (c) Experimental reflectivity map of the corresponding TIPS-pentacene microcavity in the strong-coupling regime and a coupled-oscillator fit based on a standard three-exciton model (dashed lines). Black solid lines represent the reflectivity spectra at 10° obtained from coupled oscillator fit.

*3.1.2. Accounting for inhomogeneous broadening*

In order to describe the full spectral behavior of disordered molecular systems, we extend the TC Hamiltonian to explicitly account for inhomogeneous broadening. To achieve this, we consider that the inhomogeneously broadened spectrum represents a dense distribution of well-defined transitions. In the simplest approximation, we can decompose the spectrum into a series of non-overlapping, evenly spaced, identically broadened Lorentzian oscillators (Equation 3).

$$\sum_{j=0}^{n} \frac{A_j(\Gamma_x/2)^2}{(E-E_j)^2 + \left(\frac{\Gamma_x}{2}\right)^2} \simeq Abs(E); \; E_j = E_0 + j*D \tag{3}$$

In this fit the *n* 'excitons' are indexed with j, each with peak energy $E_j$, separated by spacing $D$, and with a common full-width at half-max $\Gamma_x$ to capture exciton damping. Though clearly a simplification of the complex structure arising from the rich interplay of vibronic coupling and structural disorder, this approach suffices to capture an important change in behavior in the strong-coupling regime. These excitations collectively form a manifold that exhibits even energy spacing and interacts with light in a manner that is highly constrained by each peak's contribution to the overall spectrum. Namely, for a given choice of *n*, $D$, and $\Gamma_x$, the amplitudes $A_j$ are fully determined by the measured absorption profile. Incorporating this distribution of excitations into an extended, multi-oscillator TC model, the amplitudes map onto the individual light-matter coupling strengths as $\Omega_j \propto \sqrt{A_j}$.

$$\widehat{H}_{TCM} = \begin{bmatrix} E_c(\theta) - i\Gamma_c & \Omega_1 & \Omega_2 & \ldots & \Omega_n \\ \Omega_1 & E_1 - i\Gamma_x & 0 & \ldots & 0 \\ \Omega_2 & 0 & E_2 - i\Gamma_x & \ldots & 0 \\ \vdots & \vdots & \vdots & \ddots & \vdots \\ \Omega_n & 0 & 0 & \ldots & E_n - i\Gamma_x \end{bmatrix} \tag{4}$$

Precisely as above, we can diagonalize this matrix to determine the coupled eigenstates and extract their linewidths and photonic character to compute the corresponding reflectivity dispersion.

The results of this approach as applied to crystalline P3HT are shown in Figure 4. Here, we use a series of 39 Lorentzians to describe the absorption spectrum (Figure 4a), though we emphasize that the result is insensitive to *n* provided the sum of oscillators does not significantly deviate from the experimental spectrum. The angle-dependent reflectivity of the corresponding cavity and a spectral cut at 10° are presented in Figure 4b. Despite the significantly reduced $\Omega_j$ for each oscillator compared to the three-oscillator model (Figure 3a), we quantitatively capture the same LP and UP dispersion and achieve a much more accurate description of their relative linewidths and intensities. Moreover, this model accurately reproduces the subtle intermediate modulations highlighted above (arrows). Repeating this analysis for the semi-crystalline and amorphous cavities, we find much better fidelity to the experimental dispersions across the full series (see SI Figure 4a-f), including the variation in the weak intermediate bands that is obscured by the conventional approach.

The description of the experimental reflectivity using our method rivals the other conventional approach in the field, transfer matrix modelling. The crucial advantage to our multi-oscillator model, though, is that it provides ready access to the distribution of eigenstates and an understanding of how they contribute to the collective signal, as depicted in Figure 4c for the 10° cut. For instance, we see that the prominence of the LP and UP bands arises from two different origins. Considering the depth of modulation tracks with photon character, the LP is dominated by a single, highly photonic eigenstate. But the UP instead consists of a dense series of only moderately photon-mixed eigenstates. The latter result is in accord with the state structure predicted using Redfield theory to describe disordered carbon nanotube polaritons.[38] Moreover, our model yields numerous middle states, each exhibiting nonzero photonic character and collectively contributing to the (experimentally observed) deviation of the signal from complete reflectance. This intermediate band has been described elsewhere as gray or subradiant states,[52,53] and it is a hallmark of disorder effects. As in previous, theoretically motivated explorations of the role of disorder in exciton-polariton formation, we find that the conventional binary picture between bright polaritons and intracavity dark states no longer holds. Within this manifold, it is straightforward to identify the groups of eigenstates that contribute to the faint modulations in the total reflectivity spectrum. Two sub-sets of states (green, light blue) exhibit photonic character slightly beyond the baseline level, resulting in a detectable impact on the collective signal.

In addition to characterizing the eigenstates by their photonic character, it can be instructive to project them back onto the original exciton basis. In Figure 4d we plot this projection for the four eigenstates numbered in Figure 4c. We see that even the gray states exhibit significant mixing between multiple parent exciton states, with the strongest contributions coming from those closest to resonance with the coupled mode. Interestingly, there is little difference in this regard from a state in the UP band (state 31) and those in the intermediate gray band (11 and 21). Indeed, only the highly photonic LP state exhibits comparable contributions from more than 2-4 excitonic states, and nonetheless these contributions arise solely from the low-energy edge. Given this combination of photonic character and mixing between different exciton states, the question emerges whether the stronger 2.08 eV band and the subtle fluctuation at 2.31 eV could be identified as middle polariton states. Assignment of middle polaritons in equivalent broad-absorber microcavities is inconsistent, based on ill-defined thresholds of peak visibility. We consider that our approach highlights the challenge of mapping the observed states onto the binary of polaritons versus dark or even gray states. By reframing the light-matter interaction with a quasi-continuum of excitons, we see that the product of the interaction is likewise a continuum of states, exhibiting a continuum of behaviors with many of the hallmarks of polaritons.

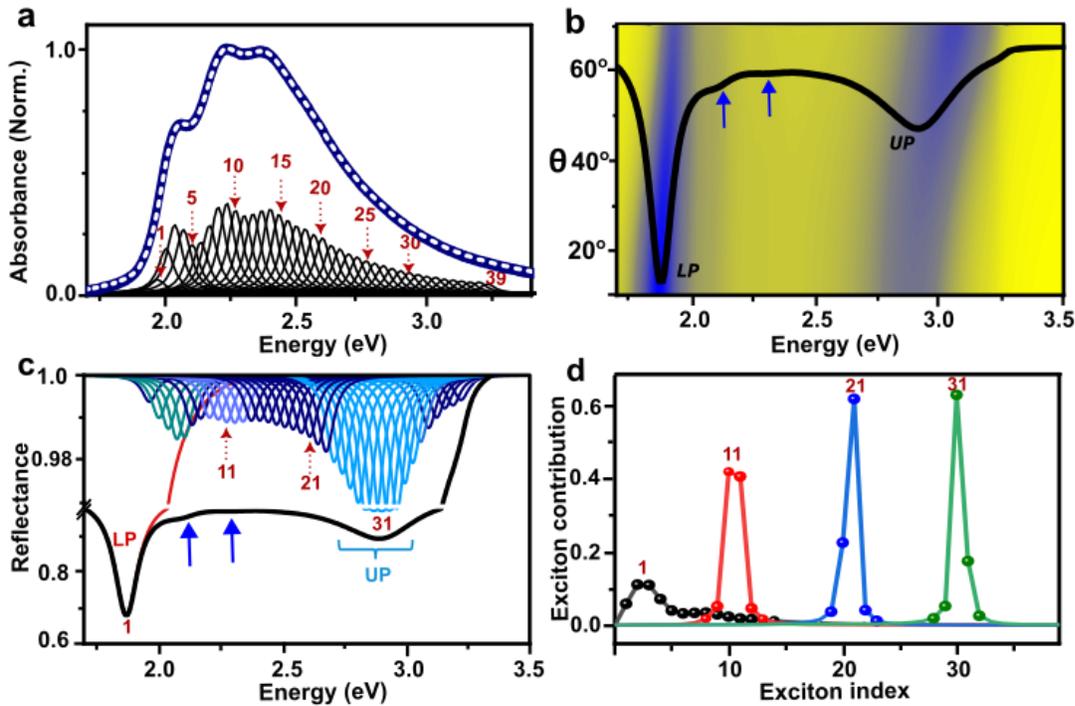

**Figure 4.** (a) Steady state absorption spectrum (solid blue line) and multi-oscillator decomposition of crystalline P3HT thin film. The white dashed line represents the sum of Lorentzian oscillators (b) Modelled angle dependent reflectivity

map of corresponding cavity overlaid with reflectivity spectrum at 10°. (c) Modelled reflectivity spectrum at 10° (black), with its eigenstate decomposition in color. The LP state is composed of a single bright eigenstate (red). The remaining spectral features result from the contribution of multiple states exhibiting less photonic character. The depth of the individual reflectance peaks provides a rough measure of the photonic character in the eigenstate. (d) Projection of the mixed eigenstates numbered 1, 11, 21, 31 in (c) onto the original exciton basis. Exciton index follows the numbering of panel (a).

### 3.2. Generality of multi-oscillator model

To highlight the applicability of our multi-oscillator approach as a general-purpose tool, we apply it to other organic semiconductor microcavities exhibiting diverse levels of disorder (Figure 5). We first consider TIPS-pentacene, which exhibits well-resolved vibronic structure in its absorption spectrum and can be described well by the three-oscillator model in Figure 3b. At the other extreme, we consider PTCDA, which has a broader absorption spectrum in the visible due to overlapping Frenkel exciton and charge-transfer excitations [54,55] and additional minor absorption bands in the ultraviolet (Figure 5b). Both materials exhibit clear strong light-matter coupling, with a typical ladder of distinct polaritons in TIPS-pentacene (Figure 5c) versus strong LP and UP accompanied by minor dispersive features near the other absorption bands in PTCDA (Figure 5d). Results from our multi-oscillator modelling in Figure 5e-f show excellent agreement for both molecules, capturing the positions and qualitative behavior of all observed bands. Thus, unlike the conventional model, the multi-oscillator approach can be safely applied to describe strong light-matter coupling to any type of absorber.

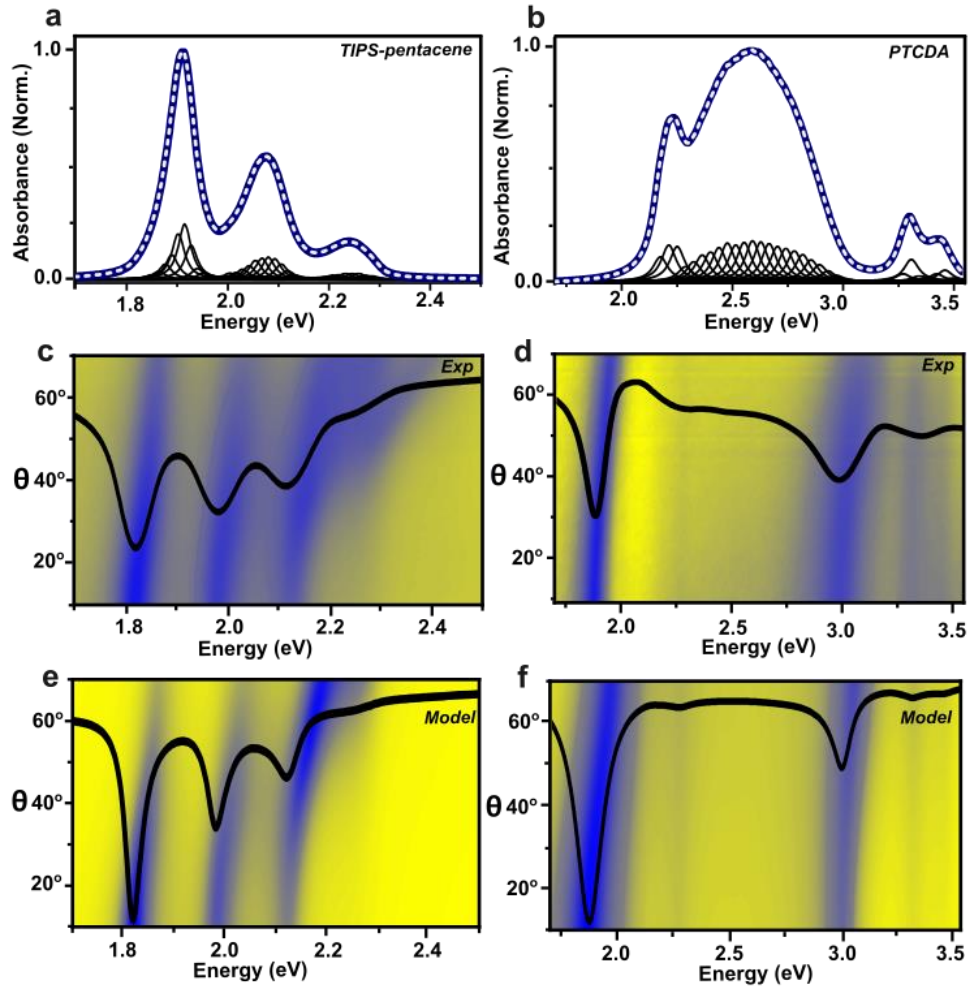

**Figure 5.** Steady state absorption spectra (solid blue line) of (a) TIPS-pentacene/PS and (b) PTCDA thin films. The absorption spectrum obtained by summing the Lorentzian oscillators used in the decomposition is represented by the white dashed lines. (c,d) Experimental reflectivity maps of corresponding microcavities overlaid with the reflectivity spectrum at 10°. (e,f) Reflectivity maps predicted from the multi-oscillator model. The black solid line shows the reflectivity spectrum at 10°.

### 3.3. Quantifying total light-matter interaction

The key parameter that defines the strong-coupling regime is the coupling constant $g$. It sets the energy scale for light-matter interactions and is suggested to be pivotal in moderating or even suppressing electron-phonon coupling in systems with prominent vibronic structure.[56] In a discrete state model, $g$ is directly linked to the Rabi splitting, which can be straightforwardly read off spectroscopic data, through[23,44,50]

$$\Delta E = 2\hbar g \propto \sqrt{\frac{N}{V}} = \sqrt{C} \tag{5}$$

Already the incorporation of well-defined middle polaritons as in Figure 5c complicates the picture, and there are inconsistent practices as to whether the system is most meaningfully described in terms of the UP-LP splitting or the gaps between adjacent eigenstates.[21,41,57,58] In our picture for broad absorbers the problem is more acute, as we predict a quasi-continuum of coupled states and the separation between them is predetermined by the arbitrary spacing between oscillators used in the decomposition. What $\Delta E$, if any, best reflects the strength of light-matter interaction in the system?

Recalling the well-known link between splitting and absorber concentration in the standard model (Equation 5), we probe the relationship between peak spacing and $\sqrt{N}$. To this end, we prepared a series of seven crystalline P3HT microcavities in the multilayer structure of Figure 1c. These structures contained pure P3HT active layers centered at the field anti-node, flanked by gelatin spacers on either side to maintain comparable detuning between cavities. The absorption spectra in Figure 6a shows that the overall absorption lineshapes of cavity-free films were relatively unchanged by the processing, with the main effect being a systematic decrease in absorbance at increased spin speed. Notably, in the reflectivity spectra shown in Figure 6b, all four characteristic features (LP, UP, and intermediate bands denoted with arrows) are preserved across all data sets. We extracted the energetic separation between each pair of modes at the point of closest approach. Plotted against the square root of the spectrally integrated absorption coefficient, a proxy for $N$, our findings in Figure 6c reveal a strong linear correlation with the UP-LP gap. None of the other separations exhibit such a linear trend. That is, despite the complex state structure and the presence of a continuum of intermediate gray states, the separation between the two strongly photonic states emerges as the parameter most reflective of the overall light-matter coupling strength. Whether in P3HT or better-structured systems like TIPS-pentacene, the intermediate separations are not a meaningful metric, as the middle bands are bounded by the original exciton transitions. Earlier consideration of inhomogeneously broadened absorbers in the strong-coupling regime similarly found that the splitting is a collective interaction across the whole ensemble.[41,44] We see that this picture holds even when the system exhibits distinct exciton fine structure. Indeed, P3HT and similar systems set up a seemingly paradoxical result. The overall coupling is collective across the ensemble of absorbers, but Figure 4d reveals that even the brightest states are not fully collective in terms of exciton contributions. Instead, the exciton contributions are localized in particular energetic regions, suggesting contrary to Houdre et al.[28] that the coupling in some circumstances may pick out a subset of nearby states.

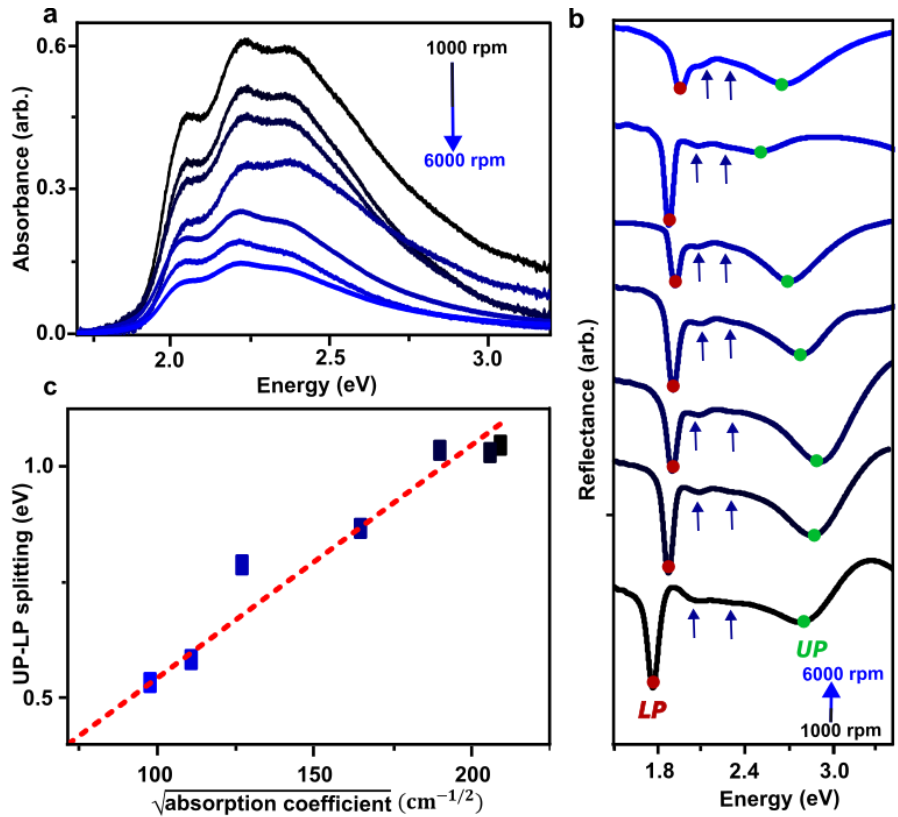

**Figure 6.** (a) Variation of crystalline P3HT absorption spectra with film thickness, tuned through spin-coating spin speed. (b) Experimental reflectivity spectra of corresponding microcavities at 10°. (c) Dependence of the measured Rabi splitting (separtion between LP and UP) on the square root of the P3HT layer absorption coefficient.

### 3.4. Origin of mid-gap modulations

To get a better picture of the origin of these intermediate grey-state features, we turn to a spectrally simpler broad absorber. The absorption spectrum Lemke dye dispersed in PMMA is presented in Figure 7a, and it is decomposed into 39 oscillators as above. In Figure 7b we show the experimental reflectivity dispersion of a Lemke dye microcavity, with the mode positions extracted from our multi-oscillator TC model superimposed as dashed lines. Given the continued good agreement with experiment, we can use our model to explore the impact of hypothetical spectral modifications. Here, we uniformly scale the absorption intensity of the three oscillators indicated in red from 200% to 25% of their original value (SI Figure 5). The resulting calculated reflectance spectra at 10° are presented in Figure 7c. For small increases or decreases, we recover the same types of subtle modulations discussed in the sections above. At the extremes, these bands approach the level of distinguishability where they would likely be classified as middle polaritons, though the precise threshold is ambiguous. This process demonstrates that the emergence of middle states doesn't necessitate introducing new excitons: precisely the same excitons are present in every calculation. Modification of the absorption lineshape simply results in a redistribution of photon character within the strong-coupling regime, leading to changes in the collective reflectivity response. This insight indicates that these subtle middle features in disordered systems reflect the underlying structural characteristics of the absorption. Their presence does not necessarily signify the existence of entirely new eigenstates but serves as a herald of a broad distribution of weakly photon-admixed gray states.

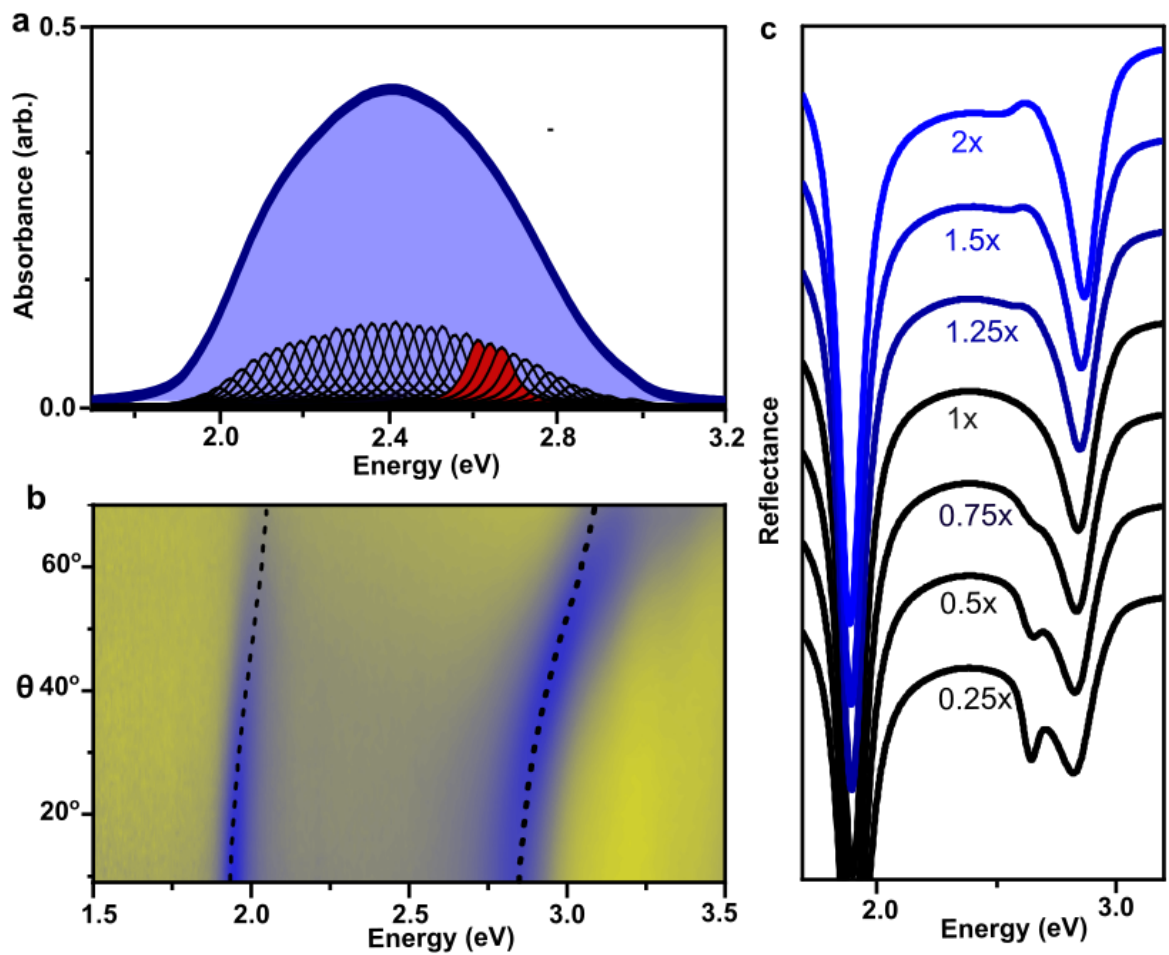

**Figure 7.** (a) Steady-state absorption spectra (solid blue line) and multi-oscillator decomposition of Lemke dye. The excitons selected for scaling are highlighted red. (b) Experimental reflectivity of a cavity a using Lemke/PMMA ratio of 0.9. The dashed line gives the modelled dispersion curves for the most photonic states. (c) Modelled reflectivity spectrum at 10° for a Lemke dye cavity with the oscillators highlighted in panel (a) scaled by the factors given.

### 3.5. Evaluation of USC

A potential pitfall of our approach is most of our model systems—P3HT, PTCDA, and Lemke dye—exhibit extreme splittings typically characteristic of the USC regime ($\Delta E_{UP-LP} > 0.2 \times E_x$ or $g > 0.1 \times E_x$).[16,49–51,59,60] In this regime, the light-matter coupling is so significant that the rotating wave approximation breaks down.[59] This necessitates the incorporation of anti-resonant and diamagnetic terms, for instance as in the Hopfield Hamiltonian shown below.[44,49,50,61,62] And yet, we have thus far described our experimental results with high fidelity using an extended TC model that lacks these terms. This fact raises questions about how to achieve and characterize USC, which will be important to apply broad-absorbing organic materials for phenomena such as superradiant phase transitions,[63] polaritonic modification of the ground state,[61,64] photon blockades,[65,66] and ultra-efficient emission.[67,68] In the following, we compare our multi-oscillator TC model to several other model approaches and experimental data on Lemke dye ($\Delta E_{UP-LP} \sim 1.25\ eV$) to evaluate the role of USC.

Similar to the TC Hamiltonian, the Hopfield Hamiltonian can be diagonalized and extended to account for homogenous broadening as follows:

$$\hat{H}_{HFS} = \begin{bmatrix} E_c(\theta) - i\Gamma_c + 2D & -i\Omega & -2D & -i\Omega \\ i\Omega & E_x - i\Gamma_x & -i\Omega & 0 \\ 2D & -i\Omega & -E_c(\theta) + i\Gamma_c - 2D & -i\Omega \\ -i\Omega & 0 & i\Omega & -E_x + i\Gamma_x \end{bmatrix} \quad (6)$$

where the diamagnetic term $D = \frac{\Omega^2}{E_x} = \frac{\hbar g^2}{E_x}$. In this form the solutions can be readily translated into reflectivity dispersions as above, and the Hamiltonian can be naturally extended to $n$ excitations in a similar fashion to TC (see SI section 6 for more details). In addition to the conventional single-exciton Hopfield Hamiltonian (HFS) shown in Equation 6, we further analyzed the experimental dispersion of Lemke dye using three other models: the standard single-exciton TC Hamiltonian (TCS), our multi-exciton TC model (TCM), and the equivalent multi-exciton Hopfield Hamiltonian (HFM). The dispersions are compared in Figure 8a to the experimental data, and evidently all models provide a reasonable description of the measurement. In other words, the introduction of additional terms in the Hopfield Hamiltonian does not necessarily improve the description of the data and may not be fully justified in this and similar systems, despite the extremely large UP-LP splitting. One important difference between the parameters used in the models is the parent photon mode position required to get a good description (solid lines). Each model requires a slightly different cavity thickness as input to produce the best fit, and ideally this would provide a means to distinguish the most appropriate description. However most broad molecular absorbers studied for USC are solution processed, and typical thickness variations within and between nominally identical samples give uncertainty on the same scale as the variations between models.[69] In the case of PTCDA, an equivalent analysis suggests the TCM model is more appropriate. Thus, the unique physics of USC does not need to be invoked to accurately describe the dispersions for the materials presented here.

We further evaluate the suitability of the USC description through other standard metrics. Fundamentally, the threshold for USC develops on the magnitude of the coupling parameter $g$. However, the value of $g$ is inherently dependent on how the exciton is defined. In the case of a single-exciton model (TCS or HFS), our Lemke dye cavity with the largest splitting yields $g \sim 0.6\ eV$, or 25% of $E_x$. And yet, when we instead describe the absorption as a disordered ensemble using TCM or HFM, the maximum value of $g$ is reduced to $\sim 0.2\ eV$, distinctly below the 10% threshold. The threshold for USC has likewise been identified based on how the UP-LP splitting varies as a function of normalized coupling constant $g' = \frac{2\hbar g}{\hbar E_x}$. In the single-exciton picture, the UP and LP energies can be described analytically by the eigenvalues of the Hopfield Hamiltonian, obtained from the solutions of Equation 7.[50]

$$(E_c^2 - E^2)(E_x^2 - E^2) = g'E_x^2 E_c^2 \quad (7)$$

In the regular strong-coupling regime, the eigenvalues are approximately linear in coupling strength $g'$, as the anti-resonant terms are negligible.

$$E_\pm \cong E_x\left(1 \pm \frac{g'}{2}\right) \quad (8)$$

In the USC regime, the anti-resonant terms cause a clear deviation from linearity in $g'$ at zero detuning.

$$E_\pm \cong E_x\left(\sqrt{1 + \frac{g'^2}{4}} \pm \frac{g'}{2}\right) \quad (9)$$

Previously, Equations (8) and (9) have been used as a test for USC, with the onset of non-linearity in $g'$ identified at a coupling ratio $\frac{2\Omega}{\omega_0} \sim 0.25$.[50,60] We employed a similar analysis for Lemke dye microcavities, varying the dye/PMMA weight ratios from 0.11 to 0.9. In Figure 8b we plot the UP and LP energies extracted at the point of minimum separation and normalized to the bare exciton energy (filled circles), against the coupling strength $g'$. The results do not follow Equation (8) (black dashed) or Equation (9) (red dashed). Nor do they agree with the outputs of our TCS model (X's), though we obtain an excellent match with our TCM model. This result indicates that these simple formulas based on a single-exciton model – the principal way that USC is treated to date – do not properly apply to broad absorbers encompassing many different transitions.

Based on these collected observations, we question the determination of USC in organic systems with broad absorption. The full data can be best described with explicit consideration of inhomogeneous broadening, but in this case none of the standard hallmarks of USC are met despite the extreme splitting and the collective nature of the interaction (Figure 6). This is not only a semantic point. The implications of USC on phenomena like modification of the ground state require a large $g$ value, but this is rarely achieved in these broad absorbers. There thus appears to be a fundamental difference in the behavior of broad and narrow absorbers in the strong-coupling regime, and efforts to explore the impact of USC may be best spent on the latter systems.

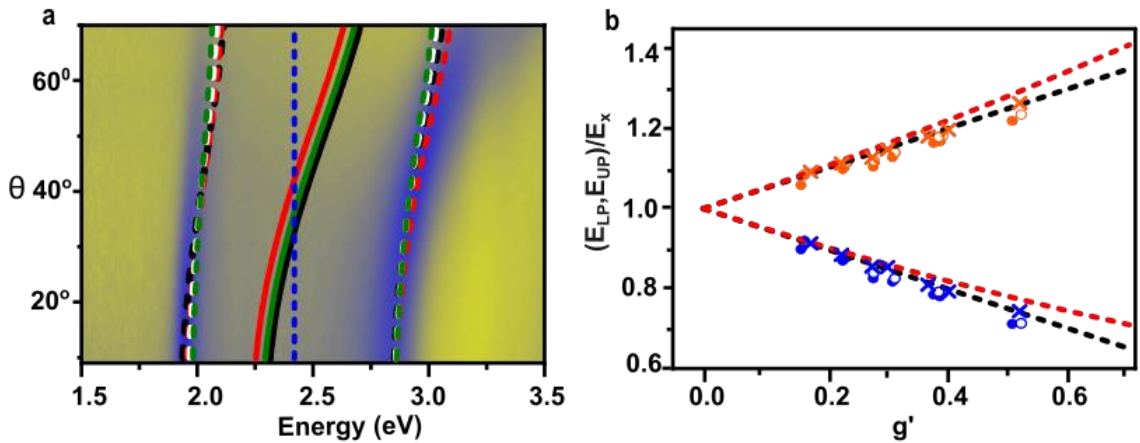

**Figure 8.** (a) Experimental angle dependent reflectivity map of cavity made of Lemke dye. Dashed lines report to UP and LP dispersion relations derived from our different models: single-exciton TC (TCS, black), multi-exciton TC (TCM, white), single-exciton Hopfield Hamiltonian (HFS, red), and multi-exciton Hopfield Hamiltonian (HFM, green). The solid lines show the corresponding bare cavity mode for each model, and the blue dashed line at 2.31 eV denotes the Lemke dye absorption peak. (b) UP (red) and LP (blue) energies, normalized to the exciton energy. Experimental data for each cavity is given as filled dots, results from the TCM model are given as open dots, and results from the TCS model are marked with X's. Dashed lines show the analytical solutions for the TCS (black, Equation 8) and HFS (red, Equation 9) models.

## 4. Conclusion

Our results highlight the importance of explicitly accounting for inhomogeneous broadening in molecular strong coupling. Using the conjugated polymer P3HT and other well-known broad absorbers, we demonstrate that manipulation of the absorption linewidths leads to subtle modulations of the optical response between the UP and LP. These widespread but overlooked features provide important insight into the nature of the states in the coupled system. They demonstrate redistribution of photon character over a manifold of intracavity gray states, which we can successfully describe by incorporating a simple approximation of electronic disorder into an extended TC model. This perspective provides an alternative description to the USC regime. Though disordered systems like these readily achieve extremely large UP-LP splittings,[16,44,51,62] the physics of USC is not required to quantitatively describe their behavior. It is instead a simple consequence of the wide energetic spread of states participating in the coupling. Crucially, our model reveals that in strong-coupled systems based on broad absorbers, the polaritons are not energetically isolated but embedded in a quasi-continuum of variably mixed states. Though the primary energetic splitting from light-matter coupling remains a collective effect, the participation in individual polaritonic eigenstates is distinctly energetically localized. Thus, the polaritons effectively pick out a subset of excitonic states rather than spreading over the full ensemble, with significant implications for polaritonic chemistry and relaxation. The presence of the dense manifold of variably photon-admixed gray states is predicted to have a major impact on polariton photophysical processes like relaxation and energy transport,[30,37,38] yet these states and the associated dynamics are poorly understood. Our simple, experimentally based approach offers a straightforward tool to describe these states and begin to quantify their role.

## Supporting Information

Supporting Information is available from the Wiley Online Library or from the author.

## Acknowledgements

This work was supported by the U.S. Department of Energy, Office of Science, Basic Energy Sciences, CPIMS Program under Early Career Research Program (Award No. DE-SC0021941), A.J.M. acknowledges the donors of the American Chemical Society Petroleum Research Fund for partial support of this research.

## Conflict of Interest

The authors declare no conflict of interest.

## Data Availability Statement

The data that support the findings of this study are available from the corresponding author upon reasonable request.